\documentclass[journal,comsoc,10pt]{IEEEtran}

\usepackage{acronym}
\usepackage{amsfonts}
\usepackage[dvips]{graphicx}
\usepackage{times}
\usepackage{cite}
\usepackage{amsmath}
\usepackage{array}
\usepackage{amssymb}
\usepackage{stfloats}
\usepackage{diagbox}
\usepackage{graphicx}
\usepackage{footnote}
\usepackage{amsthm}
\usepackage{booktabs}
\usepackage{array}
\usepackage[ruled,vlined]{algorithm2e}
\usepackage{subeqnarray}
\usepackage{cases}
\usepackage{threeparttable}
\usepackage{color}
\usepackage{epstopdf}
\usepackage{multirow}
\usepackage{tabularx}
\usepackage{enumerate}
\usepackage{multicol}
\usepackage{hyperref}
\usepackage{subfig}
\usepackage{caption}

\def\bb0{{\mathbb{0}}}

\def\bb{{\mathbf{b}}}

\def\bh{{\mathbf{h}}}

\def\bn{{\mathbf{n}}}

\def\br{{\mathbf{r}}}

\def\bx{{\mathbf{x}}}
\def\by{{\mathbf{y}}}
\def\bz{{\mathbf{z}}}
\def\b0{{\mathbf{0}}}

\def\bH{{\mathbf{H}}}
\def\bI{{\mathbf{I}}}

\def\bW{{\mathbf{W}}}

\def\sf0{{\mathsf{0}}}

\def\rm0{{\mathrm{0}}}

\acrodef{CSI}[CSI]{channel state information}
\acrodef{CSIT}[CSIT]{channel state information at the transmitter}
\acrodef{CSIR}[CSIR]{channel state information at the receiver}
\acrodef{MIMO}[MIMO]{multiple-input multiple-output}
\acrodef{SISO}[SISO]{single-input single-output}
\acrodef{MISO}[MISO]{multiple-input single-output}
\acrodef{SIMO}[SIMO]{single-input multiple-output}
\acrodef{ADCs}[ADCs]{analog-to-digital convertors}
\acrodef{SNR}[SNR]{signal-to-noise ratio}
\acrodef{AWGN}[AWGN]{additive white Gaussian noise}
\acrodef{MRT}[MRT]{maximal ratio transmission}
\acrodef{DFT}[DFT]{Discrete Fourier Transform}
\acrodef{ULA}[ULA]{uniform linear array}
\acrodef{UPA}[UPA]{uniform planar array}
\acrodef{LS}[LS]{least squares}
\acrodef{ALMMSE}[ALMMSE]{approximate linear minimum mean squared error}
\acrodef{QIHT}[QIHT]{quantized iterative hard thresholding}
\acrodef{QIST}[QIST]{quantized iterative soft thresholding}
\acrodef{SVD}[SVD]{singular value decomposition}

\def\rm#1{\mathrm{#1}}
\def\sf#1{\mathsf{#1}}
\usepackage[font=small,labelsep=none]{caption}

\begin{document}

\title{Deep Learning-based Channel Estimation for Beamspace mmWave Massive
MIMO Systems}

\author{Hengtao He, Chao-Kai Wen, Shi Jin,~and Geoffrey Ye Li
\thanks{ Manuscript received February 04, 2018; revised March 25, 2018; accepted
April 23, 2018. Date of publication XX
XX, 2018; date of current version XX XX, 2018. The work of S. Jin was supported by the National Science
Foundation (NSFC) for Distinguished Young Scholars of China with Grant
61625106 and the NSFC with Grant 61531011. The work of C.-K. Wen was supported by the Ministry of Science
and Technology of Taiwan under Grants MOST 106-2221-E-110-019 and the ITRI in Hsinchu, Taiwan. The associate editor coordinating the review of this paper and approving it for publication was Y. Gao. \emph{(Corresponding author: Shi Jin.)}}
\thanks{H.~He and S.~Jin are with the National Mobile Communications Research
Laboratory, Southeast University, Nanjing 210096, China (e-mail: hehengtao@seu.edu.cn, and jinshi@seu.edu.cn).}
\thanks{C.-K.~Wen is with the Institute of Communications Engineering, National
Sun Yat-sen University, Kaohsiung 804, Taiwan (e-mail: chaokai.wen@mail.nsysu.edu.tw).}
\thanks{G.~Y.~Li is with the School of Electrical and Computer Engineering,
Georgia Institute of Technology, Atlanta, GA 30332 USA (e-mail:
liye@ece.gatech.edu).}
}

\markboth{IEEE WIRELESS COMMUNICATIONS LETTERS. VOL. XX, NO. XX, XXX 2018}%
{Accepted paper}

\maketitle
\begin{abstract}
Channel estimation is very challenging when the receiver is equipped with a limited number of radio-frequency (RF) chains in beamspace millimeter-wave (mmWave) massive multiple-input and multiple-output systems. To solve this problem, we exploit a learned denoising-based approximate message passing (LDAMP) network. This neural network can learn channel structure and estimate channel from a large number of training data.
Furthermore, we provide an analytical framework on the asymptotic performance of the channel estimator. Based on our analysis and simulation results, the LDAMP neural network significantly outperforms state-of-the-art compressed sensing-based algorithms even when the receiver is equipped with a small number of RF chains.

\end{abstract}
\vspace{-0.5cm}

\section{Introduction}\label{sec_Introduction}

Millimeter-wave (mmWave) massive multiple-input and multiple-output (MIMO) enables the use of multi-gigahertz bandwidth and large antenna arrays to offer high data rates, which is regarded as an important technique in future wireless communications \cite{swindlehurst2014millimeter}. However, the high costs of hardware and power consumption become unaffordable when a dedicated radio-frequency (RF) chain is used for each antenna. In \cite{beamspace}, the beamspace channel model and the lens antenna array-based architecture have been proposed to reduce the number of RF chains. However, channel estimation for this beamspace mmWave massive MIMO system is extremely challenging, especially when the antenna array is large and the number of RF chains is limited.

The lens antenna array exhibits energy-focusing capability and the received signal matrix from the lens antenna array can be characterized by sparsity and concentration. Therefore, a reliable support detection (SD)-based channel estimation scheme has been proposed in \cite{SD}, which decomposes the total beamspace channel estimation problem into a series of sub-problems.
In \cite{SCAMPI}, the channel matrix is first regarded as a  $2$-dimensional ($2$D) natural image and then the cosparse analysis approximate message passing (SCAMPI) algorithm, derived from the image recovery field, has been provided. The SCAMPI algorithm models the channel as a sparse generic L-term Gaussian mixture (GM) probability distribution and uses the expectation-maximization (EM) algorithm to learn the GM parameters from current estimated data.

In this article, we present our initial results in deep learning for beamspace mmWave massive MIMO systems. We regard the channel matrix as a $2$D natural image and apply the learned denoising-based approximate message passing (LDAMP) neural network \cite{LDAMP}, which incorporates the denoising convolutional neural network (DnCNN) \cite{DnCNN} into the iterative sparse signal recovery algorithm for channel estimation. To the best of our knowledge, this study is the first to use deep learning technology for beamspace channel estimation. The LDAMP network utilizes the large number of channel matrices as training data and can be applied to a variety of selection networks. Furthermore, we provide an analytical framework on the asymptotic performance of LDAMP in channel estimation. From the analytical and simulated results, the LDAMP network outperforms the state-of-the-art compressed sensing (CS)-based algorithms and can achieve excellent performance even with a small number of RF chains.

The remaining part of this paper is organized as follows. Section \ref{System_model}
outlines the system model and formulates channel estimation as a signal recovery problem. Next, The LDAMP network and an analytical framework are provided in Section \ref{LDAMP}. Then, Numerical results are presented in Section \ref{Simulation}. Finally, Section \ref{con} concludes
the paper.
\vspace{-0.2cm}
\section{System model}\label{System_model}
In this section, we introduce the beamspace mmWave massive MIMO system and formulate the channel estimation as a signal recovery problem.

\begin{figure}
  \centering
  \includegraphics[width=7cm]{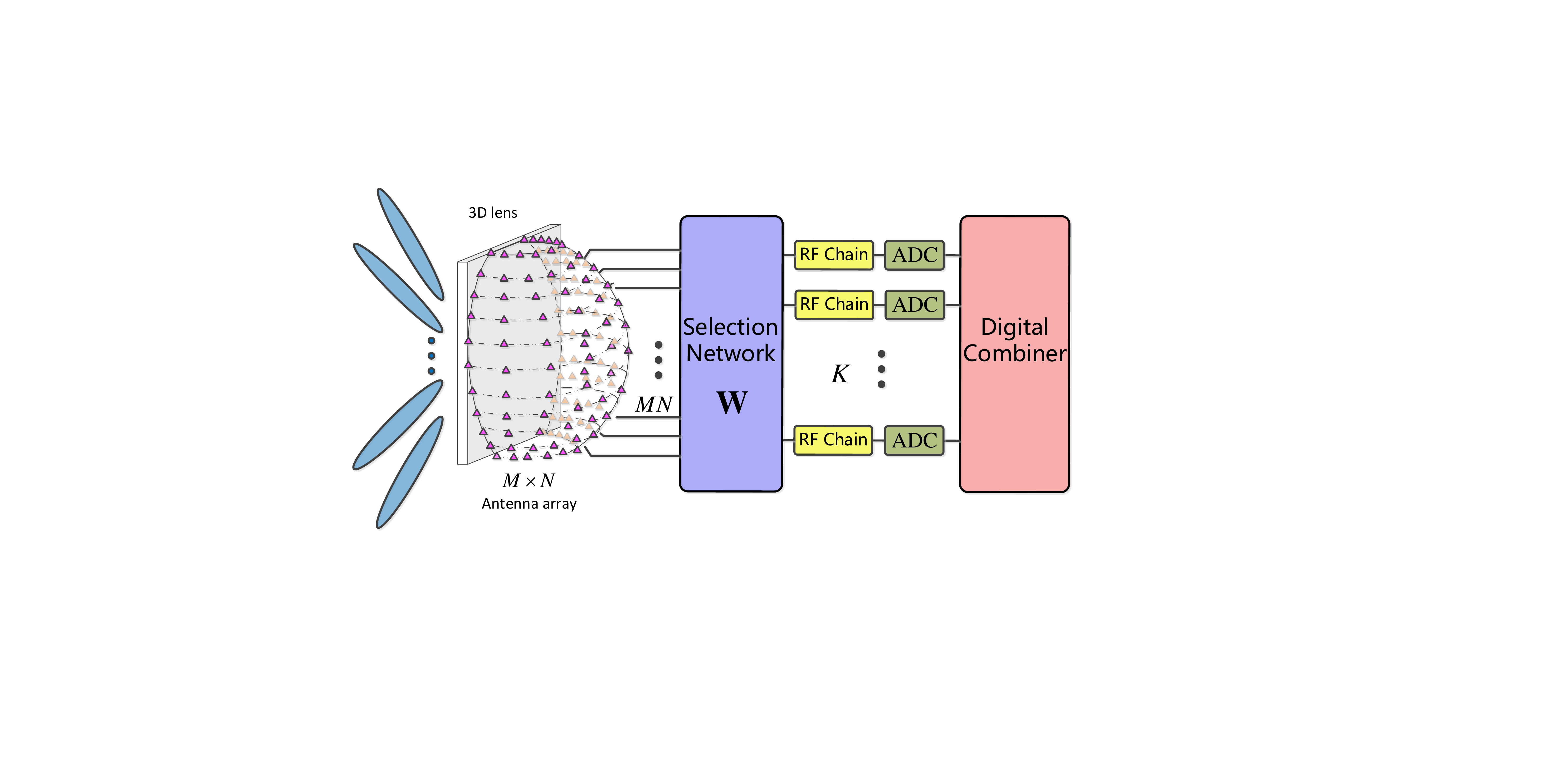}
  \caption{.~~Millimeter-Wave receiver with an EM lens and
an $M \times N$ antenna array placed on the focal plane of the lens.\vspace{-0.2cm}}\label{fig1}
\end{figure}

Fig. \ref{fig1} illustrates a typical mmWave massive MIMO system where the base station (BS) has one EM $3$D lens equipped with an $M \times N$
antenna array, and the $MN$ antennas are connected to $K$ RF chains through the $K \times MN$
selection network. In order to save the cost of power consumption of RF chains,  $K \ll MN$ is considered usually.
In Fig. \ref{fig1}, the selection network is denoted by matrix $\mathbf{W} \in \mathbb{R}^{K\times MN}$, with each entry being $\pm1$. That is, fully-connected $1$-bit phase shifters are used and normalized by dividing $\sqrt{MN}$. We assume the mmWave system with one user with a single antenna for convenience even if it can be easily extended to multiple users as long as the pilot signals for different users are orthogonal in time.

We adopt the widely used Saleh-Valenzuela channel model for mmWave communications \cite{SD}. The beamspace channel matrix can be expressed as
\begin{equation}\label{eqh}
  \bH=\sqrt{\frac{MN}{P+1}}\sum\limits_{i=0}^{P}\alpha^{(i)}\mathbf{A}(\phi^{(i)},\theta^{(i)}),
\end{equation}
where $\bH \in \mathbb{R}^{M\times N}$ denotes beamspace channel matrix\footnote{In this paper, we consider simplified real-valued model as neural network always performed in real-valued domain, and is easily extended to complex channel matrix by projecting complex-valued $\bH$ into a real-valued matrix.}, $P+1$ is the number of paths, $\alpha^{(i)}$ denotes the gain of the $i$-th path, $\phi^{(i)}$ and $\theta^{(i)}$ represent the azimuth and elevation AoAs of the incident plane wave, respectively; and $\mathbf{A}(\phi^{(i)},\theta^{(i)})$ refers to the antenna array response matrix,
which is determined by its geometry.

The beamspace MIMO channel estimation can be regarded as a typical signal recovery problem. First, we obtain the beamspace channel vector $\bh \in \mathbb{R}^{M N\times 1}$ by vectorizing $\bH$ for convenience. In the uplink training phase, the user sends the training
symbol $s$ to the BS, and the received signal vector $\by \in \mathbb{R}^{MN \times 1}$ at the BS is given by
\begin{equation}\label{eqCE}
  \by=\bh s+\bn,
\end{equation}
where $\bn\sim \mathcal{N} (\mathbf{0},\sigma_n^2\bI)$ represents a Gaussian noise vector. Given a selection network $\bW$ at the receiver, the received signal $\br$ from the RF chain can be expressed as
\begin{equation}\label{eqr}
  \br=\bW\by=\bW\bh+\bar{\bn},
\end{equation}
where $\bar{\bn}=\bW\bn$ is the equivalent noise after the selection network at the receiver, which is assumed to follow $ \mathcal{N} (\mathbf{0},\sigma_n^2\bI) $ as each entry in matrix $\mathbf{W}$ is normalized by dividing $\sqrt{MN}$. Furthermore, we set $s=1$ for convenience as pilot signal is known at the receiver side.

From (\ref{eqh}), the elements of the beamspace channel vector are not independent. They are correlated through the antenna array response matrix. This characteristic is highly similar to a $2$D natural image; that is, the channel is sparse, and the
changes between adjacent elements are subtle. Therefore, the LDAMP network, originated from image recovery, can be applied to exploit the correlation in beamspace channel estimation.
\vspace{-0.2cm}
\section{LDAMP-based channel estimation}\label{LDAMP}
In this section, we present the LDAMP-based method to estimate the beamspace channel vector $\bh$ from the received signal $\br$ and the given selection network $\bW$ in (\ref{eqr}). Besides, an asymptotic performance of the LDAMP network is investigated based on state evolution (SE) analysis.
\subsection{LDAMP network}
 Deep learning has been recently applied to solve compressive image recovery \cite{LDAMP} and wireless communications \cite{ML2OFDM,DL2017wang}.
Inspired by \cite{LDAMP}, we propose a deep learning-based channel estimation method, named LDAMP network.

\begin{figure}[h]
  \centering
  \includegraphics[width=7cm]{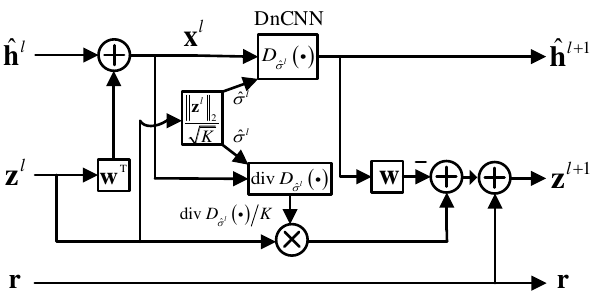}
  \caption{.~~The $l$-th layer architecture of the LDAMP network.\vspace{-0.2cm}}\label{fig2}
\end{figure}

The LDAMP neural network consists of $L$ layers connected by cascade way. Each layer has the same structure. As illustrated in Fig. \ref{fig2}, each layer of the LDAMP network contains the same denoiser $D_{\hat{\sigma}^l}(\cdot)$, a divergence estimator $\mathrm{ div} D_{\hat{\sigma}^l}(\cdot)$, and tied weights. The denoiser $D_{\hat{\sigma}^l}(\cdot)$ performed by the DnCNN is used to update $\bh$.
\vspace{-0.2cm}


\begin{figure*}[t]
  \centering
  \includegraphics[width=13cm]{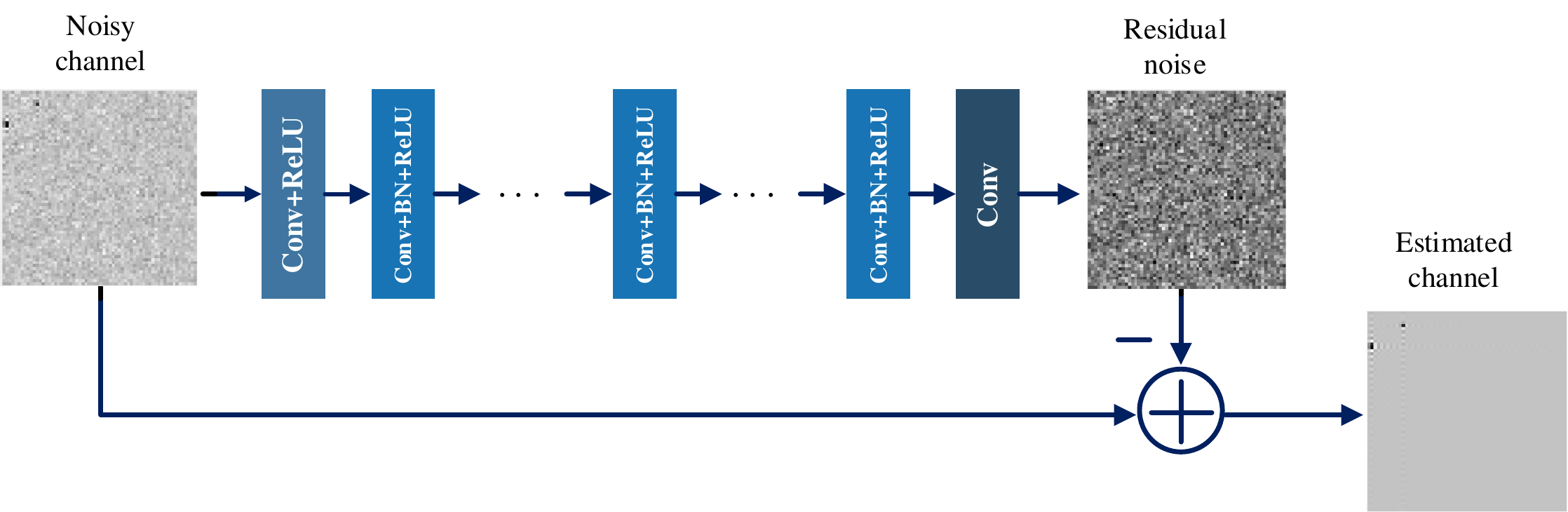}
  \caption{.~~Network architecture of the DnCNN denoiser.\vspace{-0.5cm}}\label{fig3}
\end{figure*}

Before introducing the DnCNN denoiser, we provide a brief description for the LDAMP network.
For the $l$-th layer of the LDAMP neural network, channel is estimated as follows
 \begin{align}
\bz^{l+1}&= \br-\mathbf{W}\hat{\bh}^{l+1}+\frac{1}{K}\bz^{l}  \mathrm{div} D_{\hat{\sigma}^l}(\hat{\bh}^{l}+\mathbf{W}^T\bz^{l}),\label{noise1} \\
\hat{\bh}^{l+1} &= D_{\hat{\sigma}^l}(\hat{\bh}^l+\mathbf{W}^T\bz^l),\label{noise}
\end{align}
where  $\hat{\bh}^{l}$ is the $l$-th layer input of the channel, $\bz^{l}$ represents the $l$-th layer input of the residual vector, and $\hat{\sigma}^l$ denotes a parameter of the denoiser, which is defined as $\hat{\sigma}^l = \|\bz^l\|_2/\sqrt{K}$.

  If we regard the input of the denoiser, $\bx^{l}=\hat{\bh}^l+\mathbf{W}^T\bz^l$ as a noisy channel vector
  \begin{equation}\label{eqAWGN}
    \bx^{l} =\bh+ \hat{\bn}^{l},
  \end{equation}
 then the equivalent noise $\hat{\bn}^{l}= \hat{\bh}^l-\bh + \mathbf{W}^T\bz^l
 \sim \mathcal{N}(\mathbf{0},(\hat{\sigma}^l)^2\bI)$.
 The function of denoiser $D_{\hat{\sigma}^l}(\cdot)$ is to estimate the channel $\bh$ from noisy channel $\bx^{l}$ by  removing equivalent noise $\hat{\bn}^{l}$. The equivalent noise variance $(\hat{\sigma}^l)^2$ depends on the estimated residual vector $\bz^{l}$. With increasing the number of layers, equivalent noise variance $(\hat{\sigma}^l)^2$  decreases and finally converges to a limit. Furthermore, the Onsager correction term \cite{DAMP} $\bz^{l} \mathrm{div} D_{\hat{\sigma}^{l}}(\bh^{l}+\mathbf{W}^T\bz^{l})/K$,  removes the bias from the intermediate solutions, such that the equivalent noise $\hat{\bn}^l$ follows the additive white Gaussian noise (AWGN) model expected by typical image denoisers and is uncorrelated to the channel $\hat{\bh}^{l}$.

As a precise expression for $\mathrm{div} D_{\hat{\sigma}^l}(\cdot)$ is generally difficult to obtain, we use following Monte-Carlo approximation to compute the divergence $\mathrm{ div} D_{\hat{\sigma}^l}(\cdot)$.
Given a denoiser $D_{\hat{\sigma}^l}(\cdot)$, and using an independent and identically distributed (i.i.d.) random vector $\bb \sim \mathcal{N}(\mathbf{0},\bI)$, we can estimate the divergence with
{\setlength{\arraycolsep}{0.0em}
\begin{eqnarray}
\label{eqn:div_est}
\mathrm {div} D_{\hat{\sigma}^l}&=&\lim\limits_{\epsilon\rightarrow 0}\mathbb{E}_{\bb} \left\{\bb^{T}\left(\frac{D_{\hat{\sigma}^l}(\bx^{l}+\epsilon \bb)-D_{\hat{\sigma}^l}(\bx^{l})}{\epsilon}\right)\right\}\\ &\approx &  \frac{1}{\epsilon}\bb^{T}(D_{\hat{\sigma}^l}(\bx^{l}+\epsilon \bb)-D_{\hat{\sigma}^l}(\bx^{l})),
\end{eqnarray}\setlength{\arraycolsep}{5pt}} where $\epsilon$ is an extremely small number, which we set $\epsilon=\|\bx^{l}\|_\infty/1000$.

 The denoiser used in the LDAMP network plays a key role in channel estimation.
 We consider recently developed DnCNN denoiser. The DnCNN denoiser can handle Gaussian denoising problem with an unknown noise level, which is more accurate and  faster than competing techniques. Fig. \ref{fig3} illustrates the network architecture of the DnCNN denoiser. It consists of $20$ convolutional layers. The first convolutional layer uses $64$ different $3\times 3\times 1$ filters and is followed by a rectified linear unit (ReLU).
Each of the succeeding $18$ convolutional layers uses $64$ different $3\times3\times64$ filters, each followed by batch-normalization and a ReLU.
The final convolutional layer uses one separate $3\times 3\times 64$ filters to reconstruct the signal.
Instead of learning a mapping directly from a noisy image to a denoised image, learning the residual noise is beneficial.
We plot three pseudo-color images of noisy channel, residual noise, and estimated channel in Fig. \ref{fig3}. The network is given the noisy channel $\bh+\sigma \bz$ as an input and produces residual noise $\hat{\bz}$, rather than an estimated channel $\hat{\bh}$, as an output.
This method, known as residual learning \cite{residual}, renders the network to remove the highly structured natural image rather than the unstructured noise. Consequently, residual learning improves both the training times and accuracy of a network.



\subsection{SE analysis}\label{SE}
In this section, we provide an analytical framework on the performance of the LDAMP neural network. It consists of a series of SE equations that predict the performance of the network over each layer with a large-system limit ($M,N\rightarrow\infty$), which is given by
\begin{align}
&\theta^{l+1}(\bh_o, \delta, \sigma_n^2) = \frac{1}{MN}  \mathbb{E} \| D_{\sigma^l} (\bh_o + \sigma^l_e \epsilon) -\bh_o \|_2^2, \label{eqmonte} \\
&(\sigma^l_e)^2 = \frac{1}{\delta}\theta^l( \bh_o, \delta, \sigma_n^2) + \sigma_n^2,\label{eqvai}
\end{align}
where $\bh_o$ is a deterministic realization of channel $\bh$, $\delta$ represents the measurement ratio that is defined by $K/MN$,
$\theta^l( \bh_o, \delta, \sigma_n^2)$ is the average mean-square error (MSE) of the denoiser output in the $l$-th layer network,
and $\sigma_n^2$ denotes the noise variance. The SE equations in (\ref{eqmonte}) and (\ref{eqvai}) are derived from \cite{DAMP}, which provides similar method on the performance of the D-AMP algorithms with different denoisers.

In a large-system, the SE equations can be explained from the LDAMP network itself. The explanation mainly depends on the equivalent AWGN model in (\ref{eqAWGN}), in which the average MSE of the denoiser output  is computed by the Monte Carlo method in (\ref{eqmonte}). The expectation in (\ref{eqmonte}) only concerns $\epsilon\sim \mathcal{N} (\mathbf{0},\bI)$ and the equivalent noise variance $(\sigma^l_e)^2$ is computed by (\ref{eqvai}), which is derived from (\ref{noise1}). We can obtain analytical average MSE performance of the LDAMP network by recursively updating the two equations in (\ref{eqmonte}) and (\ref{eqvai}).

\section{Simulation Results}\label{Simulation}
In this section, we provide simulated and analytical results of the LDAMP network. We assume that there is one user with a four-path mmWave channel, as described in (\ref{eqh}). The length and height of the $3$D lens are both $64$, that is, $M=N=64$. The number of layers for the LDAMP network is set to be $10$.
\subsection{Implementation details}
In our simulation, the LDAMP neural network is implemented with MatCovNet, which is a toolbox for MATLAB. The training, validation, and testing sets contain $16640$, $6400$, and $10000$ samples, respectively, and are obtained from the Saleh-Valenzuela channel model in (\ref{eqh}). Furthermore, we scale the data to the $[0,1]$ range for training the network. The LDAMP network is trained using the stochastic gradient descent method and Adam optimizer. The training rate is set to be $0.001$ initially, and then dropped to $0.0001$ and $0.00001$ when the validation error stop improving. 
\subsection{Comparison with other methods}
We compare the LDAMP network with the SD algorithm \cite{SD},  the SCAMPI algorithm \cite{SCAMPI} and  the D-AMP algorithms \cite{DAMP}\footnote{D-AMP contains many algorithms depends on different denoisers, including Gauss-AMP, bilateral-AMP, NLM-AMP and BM$3$D-AMP.}. The performance metric is NMSE, defined as
\begin{equation}\label{eqNMSE}
  \mathrm{NMSE}=\mathbb{E}\{\|\hat{\mathbf{h}}-\mathbf{h}\|_{2}^2/\|\hat{\mathbf{h}}\|_{2}^2\}
\end{equation}

\begin{figure}[h]
  \centering
  \includegraphics[width=7.4cm]{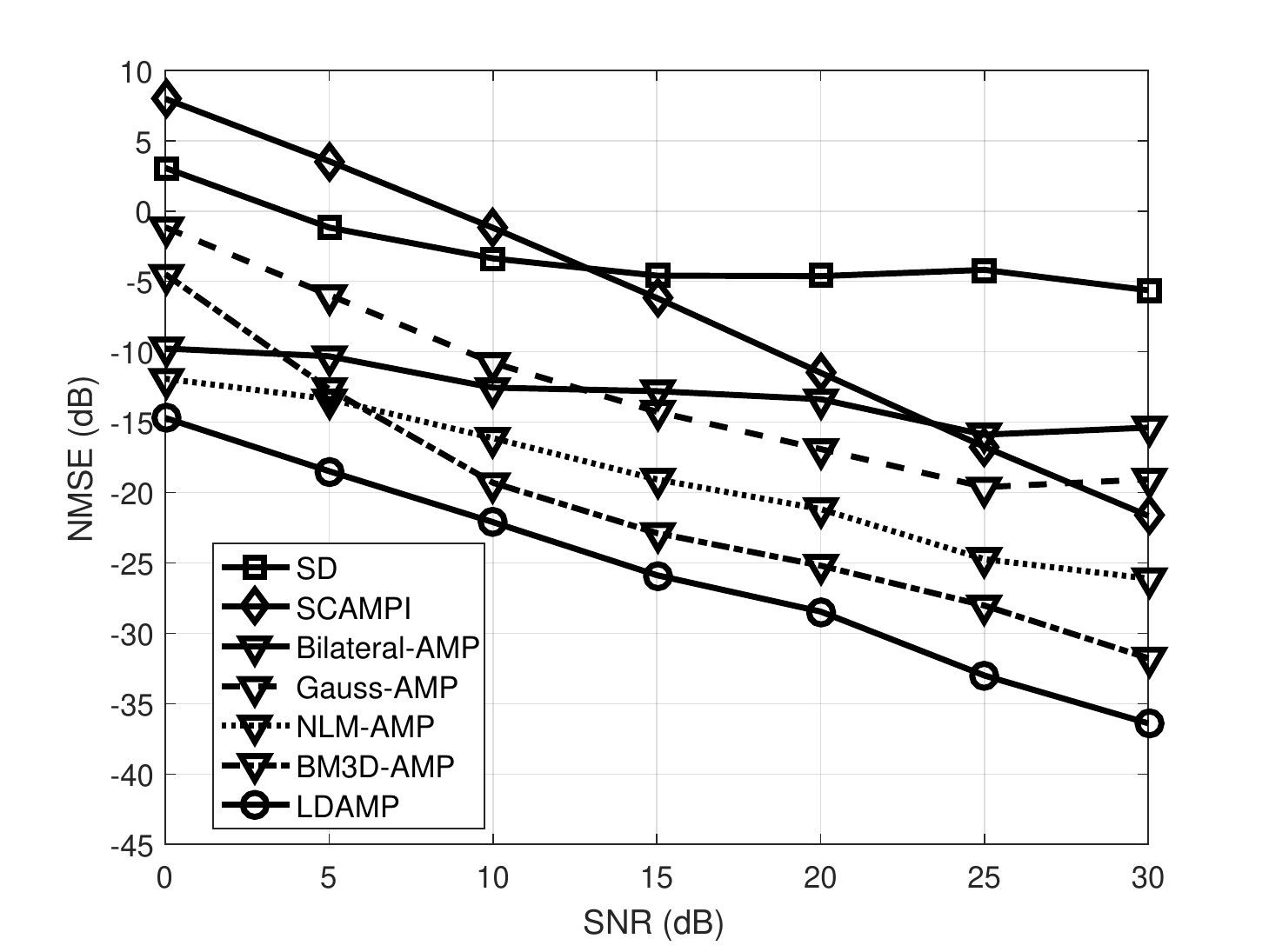}
  \caption{.~~Comparison of NMSE performance between the LDAMP network and other methods with $\delta=0.1$.\vspace{-0.2cm}}\label{fig4}
\end{figure}

Fig. \ref{fig4} compares the performance of different channel estimation methods. From the figure, the D-AMP algorithms and the LDAMP network outperform the SD and SCAMPI algorithms because of the power of the denoisers. Furthermore, the LDAMP network outperforms the start-of-the-art D-AMP algorithms even if  BM$3$D-AMP  is regarded as the most accurate algorithm for compressive signal recovery \cite{DAMP}. The excellent performance of the LDAMP network is attributed to the utilization of a large number of training data, which results in the superiority of deep learning technology.
\subsection{Analytical performance}
\begin{figure}[h]
  \centering
  \includegraphics[width=7.4cm]{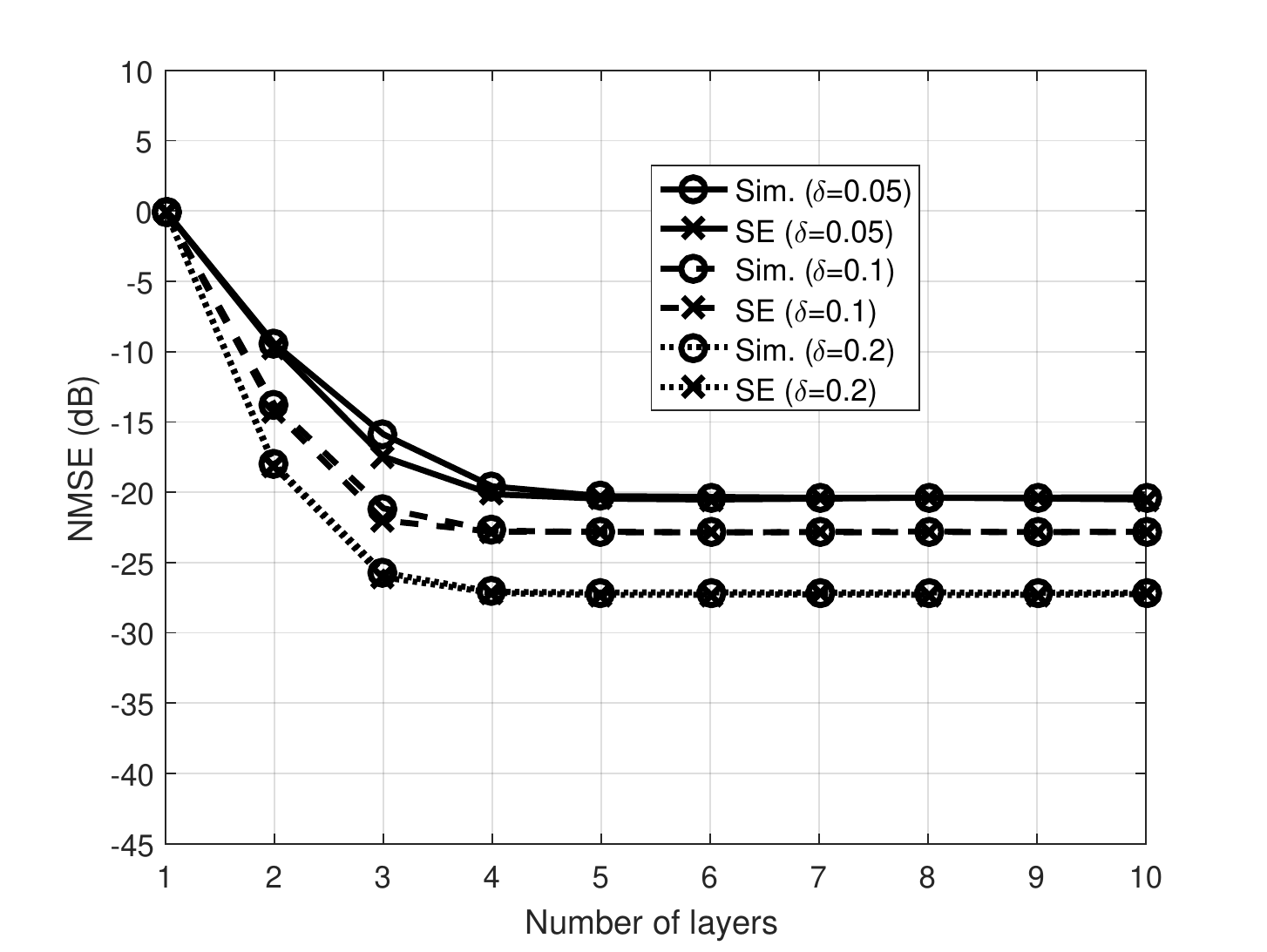}
  \caption{.~~SE analysis of LDAMP network for different measurement ratios with SNR=$10$ dB.\vspace{-0.5cm}}\label{fig5}
\end{figure}

The analytical framework on the performance of the LDAMP network for channel estimation in Section \ref{SE} is validated in Fig. \ref{fig5}. From the figure, the SE equations can precisely evaluate the performance of the LDAMP network. Therefore, instead of performing time-consuming Monte Carlo simulation to obtain NMSE, we can accurately and quickly predict the behavior by using this analytical framework. Furthermore, from Fig. \ref{fig5}, the LDAMP network converges within five layers, which demonstrates its simplicity and practicality.
\subsection{Impact of measurement ratio}
\begin{figure}[h]
  \centering
  \includegraphics[width=7.4cm]{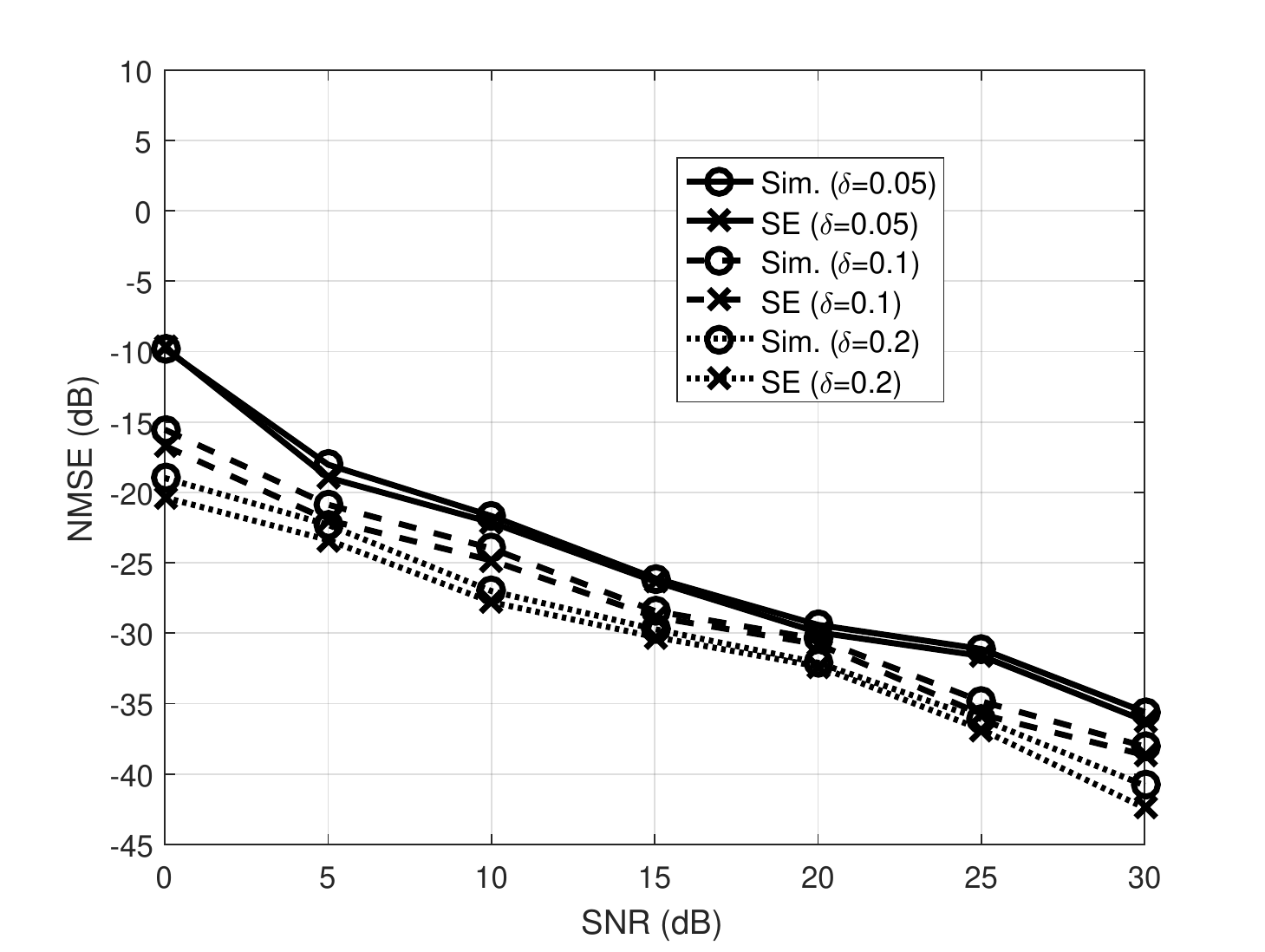}
  \caption{.~~NMSE performance of LDAMP network with different measurement ratios.\vspace{-0.2cm}}\label{fig6}
\end{figure}

As previously indicated, the measurement ratio, i.e., the number of RF chains influences the performance of LDAMP for channel estimation. Fig. \ref{fig6} illustrates NMSE with different measurement ratios. From the figure, the performance of the LDAMP network can be improved with the increase of the measurement ratio. Interestingly,  LDAMP still achieves superior performance even when measurement ratio $\delta=0.05$, thereby indicating that only a small number of RF chains are required at the receiver for channel estimation. Thus,  hardware cost and power consumption will be significantly decreased.
\section{Conclusions}\label{con}
We have developed a novel deep-learning-based channel estimation method for beamspace mmWave massive
MIMO systems. This network inherits the superiority of iterative signal recovery algorithms and deep learning technology, and thus presents excellent performance. The LDAMP network is easy to train and can be applied to a variety of selection networks. Furthermore, the LDAMP network can achieve excellent performance even with a small number of RF chains at the receiver, which validates its practicality and applicability. We have also provided an analytical framework on the performance of the LDAMP network in a large system regime, which can accurately predict performance within a short time. Our initial results have shown the potential of the deep learning method for mmWave channel estimation.

\end{document}